\pdfoutput=1
\documentclass[hyperref, journal=nalefd, manuscript=letter, layout=twocolumn]{achemso}
\usepackage{graphicx}% Include figure files
\usepackage{amsmath,amssymb}
\usepackage{xcolor}
\usepackage{pdfpages}

\title{Dynamic Control of Nonequilibrium Metal--Insulator Transitions}
\author{Joseph Kleinhenz}
\affiliation{Department of Physics, University of Michigan, Ann Arbor, Michigan 48109, USA}
\author{Igor Krivenko}
\affiliation{Department of Physics, University of Michigan, Ann Arbor, Michigan 48109, USA}
\author{Guy Cohen}
\email{gcohen@tau.ac.il}
\affiliation{School of Chemistry, Tel Aviv University, Tel Aviv 69978, Israel}
\author{Emanuel Gull}
\email{egull@umich.edu}
\affiliation{Department of Physics, University of Michigan, Ann Arbor, Michigan 48109, USA}


\graphicspath{{figures/}}
\newcommand{\doi}[1]{\href{http://dx.doi.org/#1}{\nolinkurl{#1}}}

\setkeys{acs}{doi = true}
\begin{document}

% limit: 150 words
\abstract{
We demonstrate a first order metal--insulator phase transition in the repulsive, fully frustrated, single-band Hubbard model as a function of the coupling to a fermion bath.
Time dependent manipulation of the bath coupling allows switching between metallic and insulating states both across the phase transition and within the coexistence region.
We propose a simple nanoelectronic device for experimentally realizing dynamic control of the bath coupling.
Analysis of the device characteristics shows that it can act as a two-terminal memristor.
}
\bigskip

%\emph{Introduction.}
Strongly correlated materials (SCMs) such as transition metal oxides exhibit remarkable intrinsic switching properties down to the nanoscale, making them an exciting future alternative to semiconductor technology \cite{takagi_emergent_2010}.
Hysteretic resistive switching effects driven by electric fields, currents, Joule heating, or photoexcitation have received much experimental
\cite{ruzmetov_electrical_2009, zhou_voltage-triggered_2013, pickett_sub-100_2012, chudnovskii_electroforming_1996, leroy_high-speed_2012, ha_electrical_2013, sakai_effect_2008, valmianski_origin_2018, stoliar_universal_2013, janod_resistive_2015, giorgianni_overcoming_2019, diener_how_2018, cavalleri_femtosecond_2001, cavalleri_evidence_2004, qazilbash_mott_2007, wegkamp_instantaneous_2014, liu_terahertz-field-induced_2012}
and theoretical
\cite{li_electric-field-driven_2015, li_microscopic_2017, han_nonequilibrium_2018, mazza_field-driven_2016, werner_ultrafast_2017, li_theory_2018}
attention in this context.
Applications include both transistors \cite{newns_mott_1998, stefanovich_electrical_2000, kim_mechanism_2004, ruzmetov_three-terminal_2010, nakano_collective_2012, yamada_ferroelectric_2013} and memristors \cite{driscoll_phase-transition_2009, pellegrino_multistate_2012, son_self-selective_2012, wong_bipolar_2013, ye_reversible_2018, bae_memristive_2013}.
Additionally, memristive systems based on SCMs promise to enable neuromorphic devices that mimic the behavior of biological neurons \cite{pickett_scalable_2013, strukov_missing_2008, kalcheim_non-thermal_2019, del_valle_electrically_2017, del_valle_caloritronics-based_2019}.
It is thought that such devices could offer lower power consumption and comparable---or even faster---switching timescales than traditional semiconductor electronics \cite{zhou_mott_2015, yang_oxide_2011, brockman_subnanosecond_2014}.
At the core of such devices is the physics of Mott metal--insulator transitions.

Here, we describe and solve a simple model showing that a Mott metal--insulator transition can be driven by proximity to a metallic region.
Furthermore, we propose a potential nanoscale device for realizing this effect and show that switching between the two states of the device can in principle be achieved on $\sim$100ps timescales.
The device may be operated as either a transistor, where the system is switched fully across the phase transition; or as a memristor, taking advantage of memory effects in the coexistence region.

%\emph{Model}.
We study the repulsive, fully frustrated, single-band Hubbard model on the infinite coordination number Bethe lattice, each site of which is coupled to a noninteracting fermion bath \cite{han_energy_2013, han_solution_2013, mitra_current-driven_2008}.
The Hamiltonian describing the Hubbard lattice is given by
\begin{align}
  H_{\mathrm{lattice}} &= -v \sum_{\left\langle i j\right\rangle, \sigma} c^\dagger_{i \sigma} c_{j \sigma} + U \sum_i n_{i \uparrow} n_{i \downarrow},
\end{align}
where $c^\dagger_{i \sigma} (c_{i \sigma})$ creates (annihilates) lattice fermions with spin $\sigma$ on site $i$; $v$ is the lattice hopping matrix element; and $U$ is the on-site Coulomb repulsion.
We use the hopping $v$ as our unit of energy, $\hbar / v$ as our unit of time and set $\hbar \equiv 1$.
For example, a bare bandwidth of 4~eV would set our unit of time to be 0.66~fs.
In the infinite coordination number limit considered here, this model may be solved exactly via the dynamical mean field theory (DMFT) \cite{metzner_correlated_1989, georges_hubbard_1992} and is known to exhibit a first order Mott metal--insulator transition as a function of the interaction strength $U$\cite{rozenberg_mott-hubbard_1994}.

The Hamiltonian describing the baths is given by
\begin{subequations}
\begin{align}
  H_{\mathrm{bath}} &= \sum_{i} H_{\mathrm{bath}}^{(i)}, \\
  \begin{split}
    H_{\mathrm{bath}}^{(i)}
    &= \sum_{k\sigma} \epsilon_{k} b^\dagger_{ik\sigma} b_{ik\sigma},  \\
    &+ \sum_{k\sigma} V_k(t) c^\dagger_{i \sigma} b_{ik\sigma} + V_k^*(t) b^\dagger_{ik\sigma} c_{i \sigma} .
  \end{split}
\end{align}
\end{subequations}
Here, $b^\dagger_{i k \sigma}(b_{i k \sigma})$ creates (annihilates) bath fermions coupled to site $i$ with spin $\sigma$ and quasimomentum $k$, and $V_k(t)$ is the tunneling matrix element describing hopping between the lattice and the baths.
The time dependence of the bath hopping is parameterized by a dimensionless coupling strength $\lambda(t)$ so that $V_k(t) = \lambda(t) V_k$.
The effect of the bath is characterized by a coupling density $\Gamma_{\mathrm{bath}}(\omega) = \pi \sum_k |V_k|^2 \delta(\omega - \epsilon_k)$ that parameterizes the bath dispersion $\epsilon_k$ and tunneling matrix elements $V_k$.
We choose a flat coupling density with soft-edges $\Gamma_{\mathrm{bath}}(\omega) = \Gamma / \left[\left(1 + e^{\nu \left(\omega - D\right)}\right) \left(1 + e^{-\nu \left(\omega + D\right)}\right)\right]$, with parameters $\Gamma = 1 v$, $\nu = 10v^{-1}$ and $D=4v$.
Time dependent manipulation of the bath coupling has previously been introduced as a method to induce cooling of the system \cite{werner_light-induced_2019, werner_entropy-cooled_2019}.

%\emph{Methods}.
An exact solution of the model is given by the nonequilibrium DMFT mapping \cite{aoki_nonequilibrium_2014}.
DMFT maps the lattice model to an Anderson impurity model with a self-consistently determined hybridization function $\Delta_\sigma(t, t')$ given by
\begin{align}
  \begin{split}
    \Delta_{\sigma}(t, t')
    &= v^2 G_{\sigma}(t, t')\\
    &+\lambda(t) \Delta_{\mathrm{bath}}(t, t') \lambda(t'),
  \end{split}
\end{align}
where $G_{\sigma}(t, t')$ is the impurity Green's function; $\Delta_{\mathrm{bath}}(t, t')$ is the hybridization between the lattice and bath; and $\lambda(t)$ is the time-dependent coupling strength.
The equations are solved by starting with an initial guess for $\Delta_{\sigma}(t, t')$, evaluating the impurity Green's function, and iterating.
In the coexistence region, the metallic and insulating solutions may be found by choosing a metallic or insulating initialization of the DMFT loop.

For the solution of the impurity model we use the one crossing approximation (OCA) \cite{pruschke_anderson_1989,eckstein_nonequilibrium_2010} formulated on the three-branch Keldysh--Matsubara contour.
In equilibrium the OCA is known to capture the qualitative physics of the Mott transition with reasonable accuracy \cite{vildosola_reliability_2015}.
We further validate our OCA results against numerically exact inchworm QMC\cite{cohen_taming_2015,antipov_currents_2017,dong_quantum_2017} data in the parameter regime where this is feasible (see supporting information).

The main quantity of interest is the time-dependent spectral function $A(\omega, t)$ which we calculate within an auxiliary current formalism \cite{sun_kondo_2001, cohen_greens_2014, lebanon_measuring_2001}.
We are specifically interested in the density of states at the Fermi energy, $A(\omega = 0, t)$, which we use to determine whether the system is in a metallic or insulating state.

%\emph{Results}.
\begin{figure*}[!htb]
  \includegraphics[]{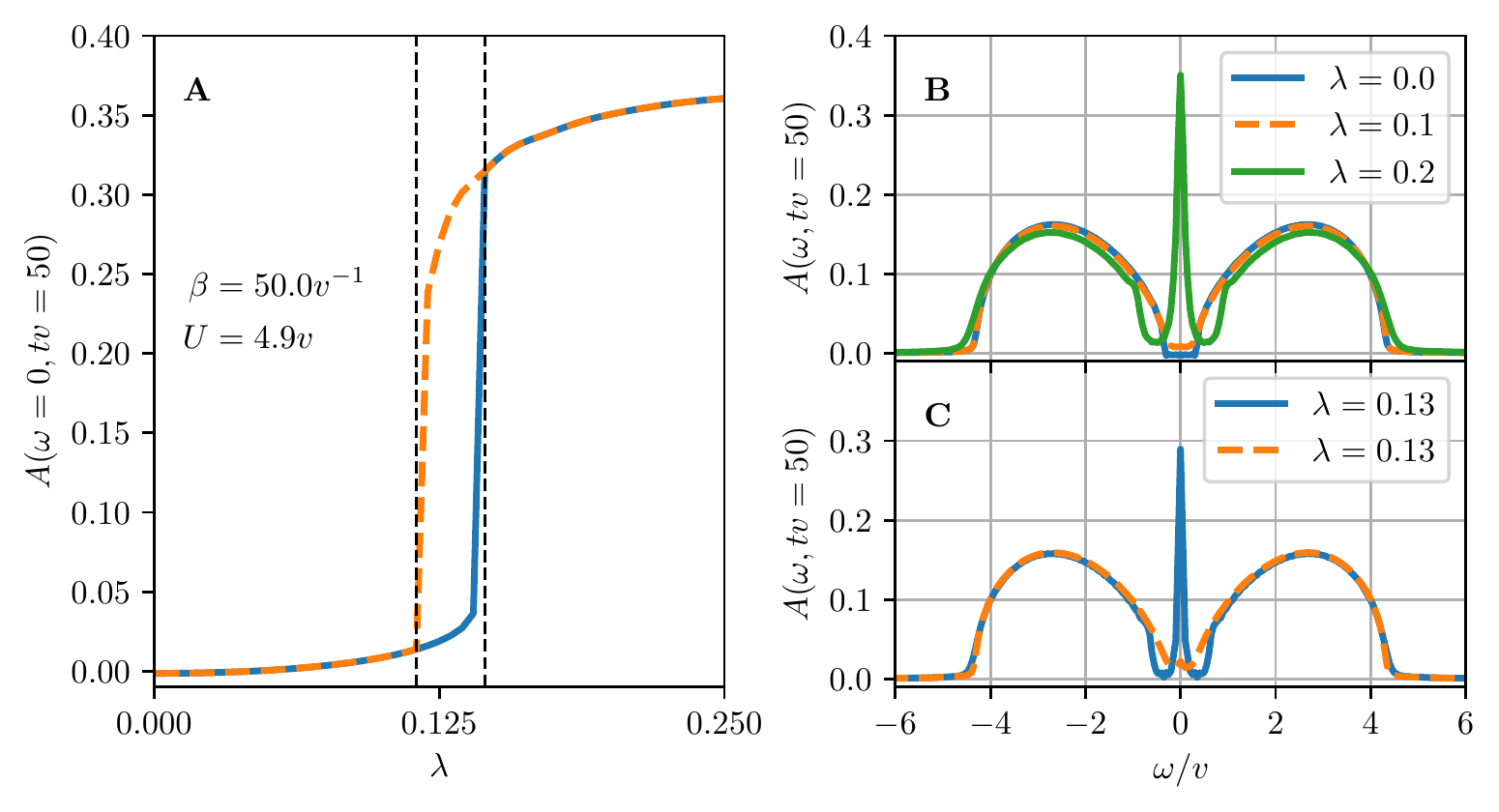}
  \caption{\label{fig:phasediagram}
  \textbf{A}: Spectral weight at $\omega = 0$ as a function of $\lambda$ for metallic (dashed orange) and insulating (solid blue) initialization of the DMFT loop.
  The dashed black lines show the boundaries of the coexistence region ($\lambda_{c_1} = 0.115$, $\lambda_{c_2} = 0.145$).
  \textbf{B}: Spectral function for several different $\lambda$.
  \textbf{C}: Spectral functions of metallic and insulating solutions in the coexistence region.
  }
\end{figure*}
Fig.~\ref{fig:phasediagram} shows the equilibrium spectral function of the system as a function of the time-independent bath coupling $\lambda$.
The interaction strength $U$ is set to $4.9 v$ and the inverse temperature $\beta$ is set to $50 v^{-1}$.
These parameters are chosen so as to generate a sizable coexistence region, and are used throughout the rest of this paper.
The maximum simulation time $t_{\mathrm{max}}$ is set to $50.0v^{-1}$, which is long enough to resolve sharp features in the spectrum.
Fig.~\ref{fig:phasediagram}\textbf{A} shows that $A(\omega = 0)$ increases by several orders of magnitude as $\lambda$ is varied from $0.0$ to $0.25$, for both metallic (dashed orange) and insulating (blue) initializations of the DMFT loop.
The system goes through a first order phase transition from an insulating state at small $\lambda$ to a metallic state at large $\lambda$.
The area between the vertical dashed black lines denotes the coexistence region, where both metallic and insulating solutions are stable, as seen from the gap between the curves representing the two initializations.
Fig.~\ref{fig:phasediagram}\textbf{B} shows the full spectral function for several different values of the bath coupling $\lambda$.
When the coupling $\lambda$ becomes large enough, metallicity is induced and a sharp quasi-particle peak forms at $\omega = 0$.
Finally, Fig.~\ref{fig:phasediagram}\textbf{C} shows the full spectral function for the metallic and insulating solutions within the coexistence region.
The two phases remain distinguishable by the presence of a sharp quasiparticle peak in the metal.

\begin{figure}[!htb]
  \includegraphics[]{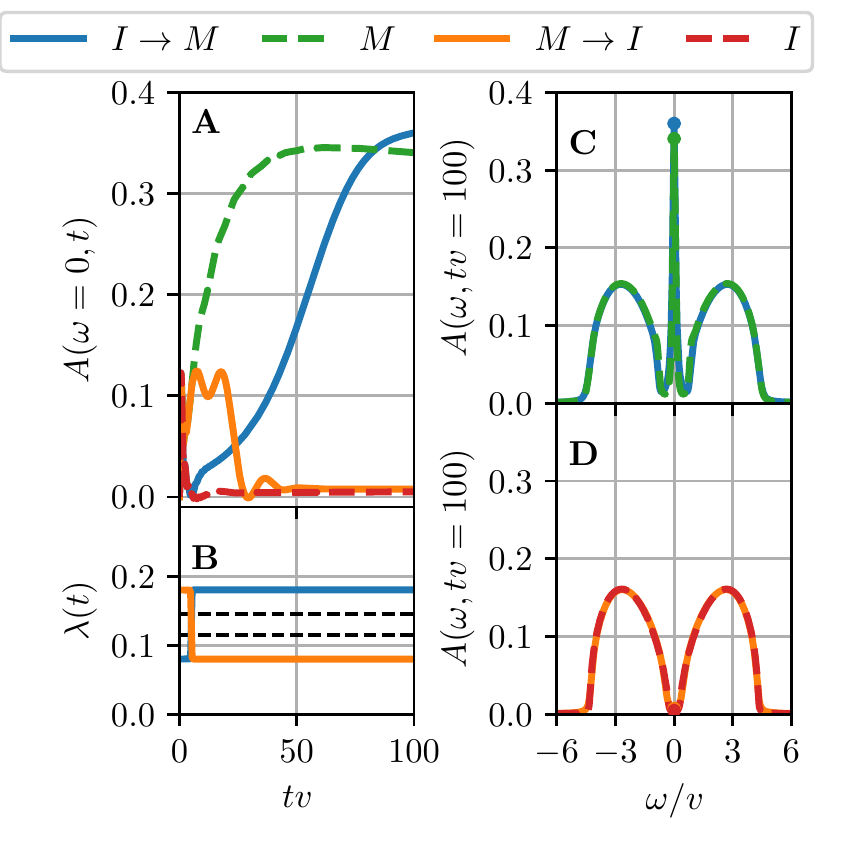}
  \caption{\label{fig:switch1}
  \textbf{A}: Evolution of $A(\omega=0, t)$ for equilibrium metallic and insulating solutions (dashed green/red) and for switched solutions (solid orange/blue).
  \textbf{B}: Switching protocol for $\lambda(t)$. Dashed black lines show the coexistence region.
  \textbf{C}: $A(\omega, tv=100)$ for equilibrium metal (dashed green) and ``switched'' metal (solid blue).
  \textbf{D}: $A(\omega, tv=100)$ for equilibrium insulator (dashed red) and ``switched'' insulator (solid orange).
  }
\end{figure}
With the equilibrium phase diagram established, we now consider two switching protocols, implemented by time dependent bath couplings $\lambda(t)$, which flip the system between the metallic and insulating phases.
In the first switching protocol, the system begins in equilibrium on one side of the phase transition.
At time $t_0$, the bath coupling $\lambda$ is rapidly quenched to a value on the opposite side.
This switching protocol is described by
\begin{align}
  \begin{split}
    \lambda(t) &= (1 - f(t)) \lambda_0 + f(t) \lambda_1, \\
    f(t) &= \frac{1}{1 + e^{-\xi\left(t - t_{0}\right)}},
  \end{split}
\end{align}
where $\xi$ sets the switching rate; $t_0$ sets the switching time; and $\lambda_0$ ($\lambda_1$) sets the initial (final) bath coupling.
Fig.~\ref{fig:switch1} shows the effect of this protocol on the system for four pairs of $(\lambda_0$, $\lambda_1$): $(\lambda_M, \lambda_M)$ (equilibrium metal), $(\lambda_I, \lambda_I)$ (equilibrium insulator), $(\lambda_M, \lambda_I)$ (``switched'' insulator), and $(\lambda_I, \lambda_M)$ (``switched'' metal).
We take $\xi = 10v$, $t_{0}v = 5$, $t_{\mathrm{max}}v = 100$, $\lambda_I = 0.08$, and $\lambda_M = 0.18$.
In Fig.~\ref{fig:switch1}\textbf{A} we plot the time evolution of $A(\omega = 0)$.
In the metal to insulator $(M \to I)$ transition, the switch rapidly destroys the metal ($A(\omega = 0)$ is suppressed).
In the insulator to metal $(I \to M)$ transition, the system gradually builds up spectral weight at $\omega = 0$ after the switch, eventually roughly matching the equilibrium metal.
Figs.~\ref{fig:switch1}\textbf{C} and \ref{fig:switch1}\textbf{D} show the full spectral function at $t_{\mathrm{max}}$ for all four realizations of the protocol.
The full spectra of the ``switched'' solutions closely resemble the corresponding equilibrium solutions, demonstrating that the protocol can switch the system between metallic and insulating states.

\begin{figure}[!htb]
  \includegraphics[]{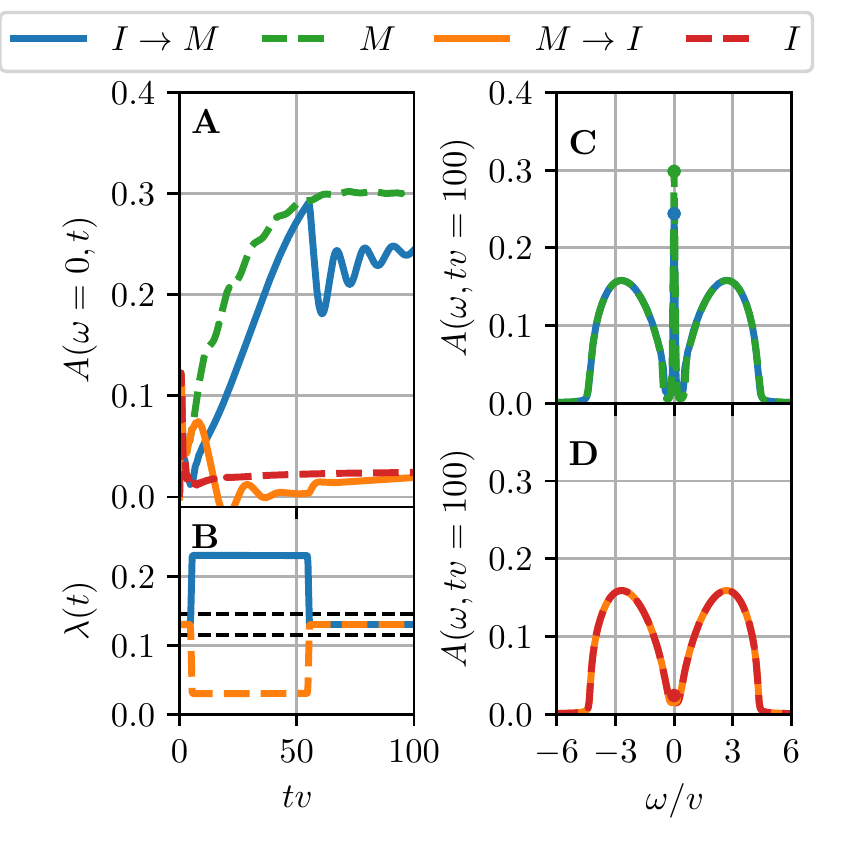}
  \caption{\label{fig:switch2}
  \textbf{A}: Evolution of $A(\omega=0, t)$ for equilibrium metallic and insulating solutions (dashed green/red) and for switched solutions (solid orange/blue).
  \textbf{B}: Switching protocol for $\lambda(t)$. Dashed black lines show the coexistence region.
  \textbf{C}: $A(\omega, tv=100)$ for equilibrium metal (dashed green) and ``switched'' metal (solid blue).
  \textbf{D}: $A(\omega, tv=100)$ for equilibrium insulator (dashed red) and ``switched'' insulator (solid orange).
  }
\end{figure}

We now consider a second switching protocol in which the system begins in equilibrium in the center of the coexistence region, in either the metallic or insulating phase.
At time $t_0$ the bath coupling is rapidly quenched to momentarily place the system outside of the coexistence region on either side of the transition; then, at time $t_1$, the bath coupling reverts to its initial (coexistence) value.
The second switching protocol is described by
\begin{align}
  \begin{split}
    \lambda(t) &= (1 - f(t)) \lambda_0 + f(t) \lambda_1, \\
    f(t) &= \frac{1}{\left(1 + e^{\xi\left(t - t_{1}\right)}\right)\left(1 + e^{-\xi \left(t - t_0\right)}\right)},
  \end{split}
\end{align}
where $\xi$ sets the switching rate; $t_0$ and $t_1$ bound the switching interval; and $\lambda_0$ and $\lambda_1$ set the initial/final and intermediate values of the bath coupling, respectively.
Fig.~\ref{fig:switch2} shows the results of this switching protocol on the system for three pairs of $(\lambda_0, \lambda_1)$: $(\lambda_c, \lambda_c)$, $(\lambda_c, \lambda_c + \Delta\lambda)$ and $(\lambda_c, \lambda_c - \Delta\lambda)$, where $\lambda_c = (\lambda_{c_1} + \lambda_{c_2})/2$ is in the center of the coexistence region, and $\Delta\lambda = 0.1$ is large enough to move the system outside of the coexistence region in either direction.
The other parameters are given by $\xi = 10v$, $t_0 v = 5$, $t_1 v = 55$, and $t_{\mathrm{max}}v = 100$.
For the equilibrium case we show both the metallic and insulating solutions.
Fig.~\ref{fig:switch2}\textbf{A} shows the time evolution of the spectral function at the Fermi energy.
In the $(M \to I)$ transition, $A(\omega = 0)$ is quickly  destroyed during the switch, and does not return when the bath coupling reverts to the coexistence region.
In the insulator to metal $(I \to M)$ transition, $A(\omega = 0)$ builds up to almost its equilibrium value during the switching period.
Afterwards, the spectral weight drops somewhat, but then recovers and appears to stabilize.
Panels \textbf{C} and \textbf{D} of Fig.~\ref{fig:switch2} show the long-time spectral function $A(\omega, t_{\mathrm{max}})$ for each of the four time evolutions.
Again, the full spectra of the ``switched'' solutions closely match the corresponding equilibrium solutions, demonstrating that the second protocol can switch the system between metallic and insulating states within the coexistence region.

We note that for both protocols the overall switching time, assuming a band width of several eV for the SCM, is on the order of $\sim$100ps.
It is important to realize that this prediction describes only the timescale needed for the electronic transitions to occur, and our minimal model does not consider any other constraints that may appear in experiments.
One should also note that this timescale is dominated by the slower transitions to the metallic state, whereas the transitions to the insulating state are substantially faster.

\begin{figure}[!htb]
  \includegraphics[width=\columnwidth]{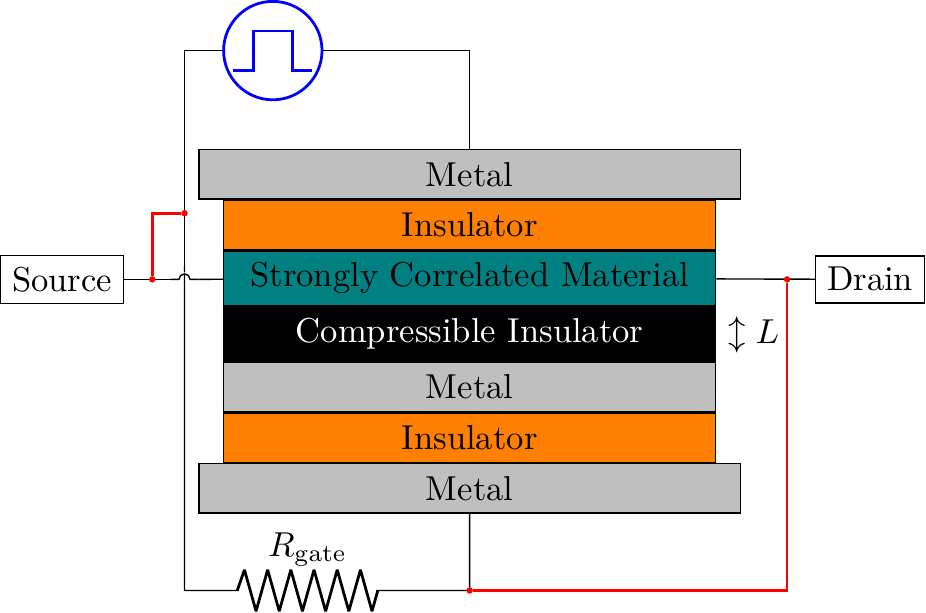}
  \caption{\label{fig:device}
  Illustration of proposed device for realizing dynamic control of $\lambda(t)$.
  The voltage between the outer metallic plates can either be externally modulated so that the device is operated like a transistor (blue) or coupled to the source-drain voltage so that the device is operated as a two terminal memristive system (red).
  }
\end{figure}
Having demonstrated the ability to dynamically control the phase of our model system through $\lambda(t)$, we shift our attention to potential experimental realizations of this effect.
Fig.~\ref{fig:device} shows an illustration of a proposed device for achieving dynamic control of $\lambda(t)$.
The core of our device consists of a SCM separated from a metal by a nanoscale, compressible, and weakly insulating region possibly composed of several polymer nanolayers.
Electronic transport across this region should be dominated by quantum tunneling effects.
This core is electrically isolated by two insulating regions and sandwiched between the plates of a capacitor (outer metallic plates).
Charging the capacitor generates a force which squeezes the compressible insulator and reduces the separation $L$ between the metal and SCM.
Since the tunneling rate $\lambda \sim e^{-L/\zeta}$ depends exponentially on the separation, we expect that (at the nanoscale) large variations in $\lambda$ can be achieved on fast timescales without the need for very large voltages or compression ratios.
This device may be operated in two modes.
In the first mode, the gate voltage across the capacitor is externally manipulated (blue signal generator in Fig.~\ref{fig:device}) to control the source-drain current via the SCM metal--insulator transition, making the device a transistor.
In the second mode, the gate voltage across the capacitor is coupled to the source--drain voltage (red connections in Fig.~\ref{fig:device}), making the device a two terminal memristor.

\begin{figure*}[!htb]
  \includegraphics[]{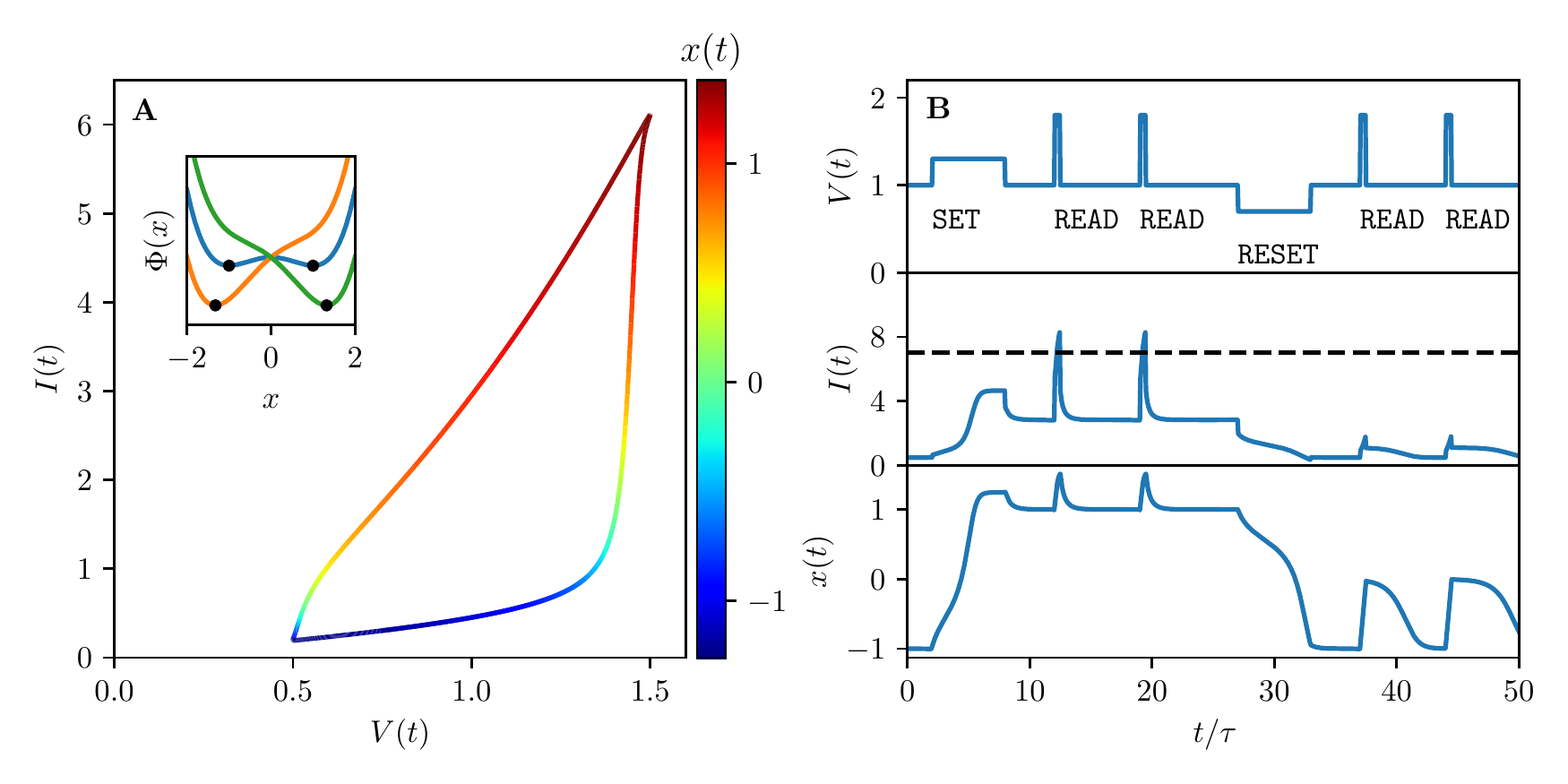}
  \caption{\label{fig:devicesim}
  Simulation of the system configured as a two terminal memristive device.
  \textbf{A}: Current voltage characteristic of device under sinusoidal driving.
  The inset shows the shape and minima of the potential $\Phi(x, \lambda)$ at $\lambda = \lambda_c, \lambda_c \pm 1$.
  \textbf{B}: Use of device as a resistive memory element.
  }
\end{figure*}
Due to computational cost, we are only able to fully simulate one switching event.
In order to further investigate and characterize the dynamics of the proposed device we consider a simple phenomenological model based on the time-dependent Ginzburg--Landau equations \cite{hohenberg_theory_1977}.
In this framework, we assume that the state of the system around the phase transition is governed by a potential
\begin{align}
  \Phi(x, \lambda) = -\left(\lambda - \lambda_c\right) x - \frac{1}{2} x^2 + \frac{1}{4} x^4,
\end{align}
where $\lambda_c = \left(\lambda_{c_1} + \lambda_{c_2}\right)/2$ is at the center of the coexistence region.
We take the order parameter $x$ to be related to the resistivity of the SCM by $R_{\mathrm{SCM}} = R_0 e^{-\alpha x}$.
At $\lambda = \lambda_c$ this potential has two stable minima at $x = \pm 1$ corresponding to metallic/insulating states with a resistivity ratio of $\exp\left({-2\alpha}\right)$.
The minimal equation of motion for $x$ is given by $\partial_t x(t) = -(1 / \tau) \partial_x \Phi\left(x, \lambda\left(t\right)\right)$ which describes exponential relaxation to equilibrium with timescale $\tau$.

We now apply this formalism to study the expected characteristics of our proposed device when configured as a memristor.
In the memristor setup, the gate voltage across the capacitor is set by the source--drain voltage $V(t)$.
We assume that the compression of the insulator is linear in the applied force so that, to leading order, the tunneling rate is given by $\lambda(t) = \gamma V^2(t) + \delta$.
Note that since $\lambda$ couples to the voltage squared, the device must be operated around a finite bias in order to have bidirectional control over $\lambda$.
For the device parameters we set $\lambda_c = 1$, $\gamma = 1$, and $\delta = 0$ so that at $V = 1$ the system is in the center of the coexistence region.
Additionally, we set $R_{\mathrm{gate}} = 10$, $R_0 = 1$, and $\alpha = 1$ so that the resistivity ratio between the insulating and conducting states is $\exp(2) \approx 7.4$.
Finally, we use $\tau = 1$ as our time unit.

Fig.~\ref{fig:devicesim}\textbf{A} shows the current--voltage characteristics (IV) of the device when driven by a sinusoidal voltage $V(t) = 1 + \frac{1}{2} \sin \left(\frac{t}{4}\right)$, where we assume that the current through the device is given by Ohm's law.
The IV forms a hysteresis loop due to the memory effect of the order parameter $x$.
In the upper part of the loop, we have $x \approx 1$; the SCM is in the metallic state; and the current is high.
In the lower part of the loop, we have $x \approx -1$; the SCM is in the insulating state; and the current is low.
Note that the hysteresis loop is not ``pinched'' (i.e. does not pass through the origin) as expected for ideal memristors \cite{chua_if_2014} because we are operating around a finite voltage bias.

Fig.~\ref{fig:devicesim}\textbf{B} demonstrates usage of the device as a resistive memory element.
The device is operated around a finite voltage bias $V_0 = 1$ so that the SCM is in the center of the coexistence region and both high and low resistivity states are stable.
The binary state of the device is encoded in the order parameter $x \approx \pm 1$.
Here the device is driven by a sequence of different pulses.
The \texttt{SET} pulse is a long low amplitude square pulse which moves the system from the $x = -1$ to the $x = 1$ state.
The \texttt{READ} pulse is a short high amplitude square pulse which causes the current to spike above a threshold (black dashed line) if the SCM is in the low resistivity $(x = 1)$ state.
The \texttt{RESET} pulse is a long low amplitude square pulse with opposite polarity to the \texttt{SET} pulse which moves the system from the $x = 1$ to the $x = -1$ state.
Combinations of these pulses allow operation of the device as a two terminal, resistive memory element.

%\emph{Conclusions}.
We have demonstrated that the repulsive, fully frustrated, single-band Hubbard model on the infinite coordination number Bethe lattice undergoes a first order metal--insulator transition as a function of a coupling to a set of free fermion baths.
By time-dependent manipulation of this coupling we are able to dynamically switch the system between its metallic and insulating states both outside and inside the coexistence region.
We propose that this effect may be realized in a nanoscale device based on manipulation of the proximity between a metal and a SCM.
Analysis of a simple model of such a device shows that it could be operated as a resistive memory element.
These results suggest a variety of directions for future work.

From a theoretical perspective, replacing the OCA impurity solver with a numerically exact method\cite{antipov_currents_2017,cohen_taming_2015,dong_quantum_2017,profumo_quantum_2015, bertrand_quantum_2019, moutenet_cancellation_2019, bertrand_reconstructing_2019} would allow us to obtain a better quantitative understanding of the timescales involved in the switching process.
It would also be of interest to investigate this type of bath-driven switching in finite dimensional models with more realistic baths and for other metal--insulator transitions, such as the transition between an anti-ferromagnetic insulator and paramagnetic metal seen in ${\mathrm{VO_2}}$.

Experimentally, we expect that a variety of ways to harness this novel switching mechanism in nanoelectronic devices and nanoscale layered materials will emerge.
Progress in this direction will rely on finding a compressible insulator with appropriate specifications, and on fabrication techniques.
While our results suggest that the fundamental limit on switching and readout time could theoretically be on the order of femtoseconds, it remains to be seen whether other limitations and engineering considerations might dominate in practical setups.
Nevertheless, the promise of being able to fabricate an efficient single-crystal memristor is certain to make the experimental challenges worth facing.

\acknowledgement{JK, EG, and IK were supported by DOE ER 46932 and by the Simons Collaboration on the Many-Electron Problem. G.C. acknowledges support by the Israel Science Foundation (Grant No. 1604/16). International collaboration was supported by Grant No. 2016087 from the
United States–Israel Binational Science Foundation (BSF).}

\begin{suppinfo}
  \begin{itemize}
    \item Comparison between the one crossing approximation and inchworm QMC.
  \end{itemize}
\end{suppinfo}

\bibliography{refs}

% need to do this to keep hyperlinks in main document
%see https://tex.stackexchange.com/questions/14620/how-to-preserve-hyperlinks-in-included-pdf
\clearpage
\includepdf[pages=-]{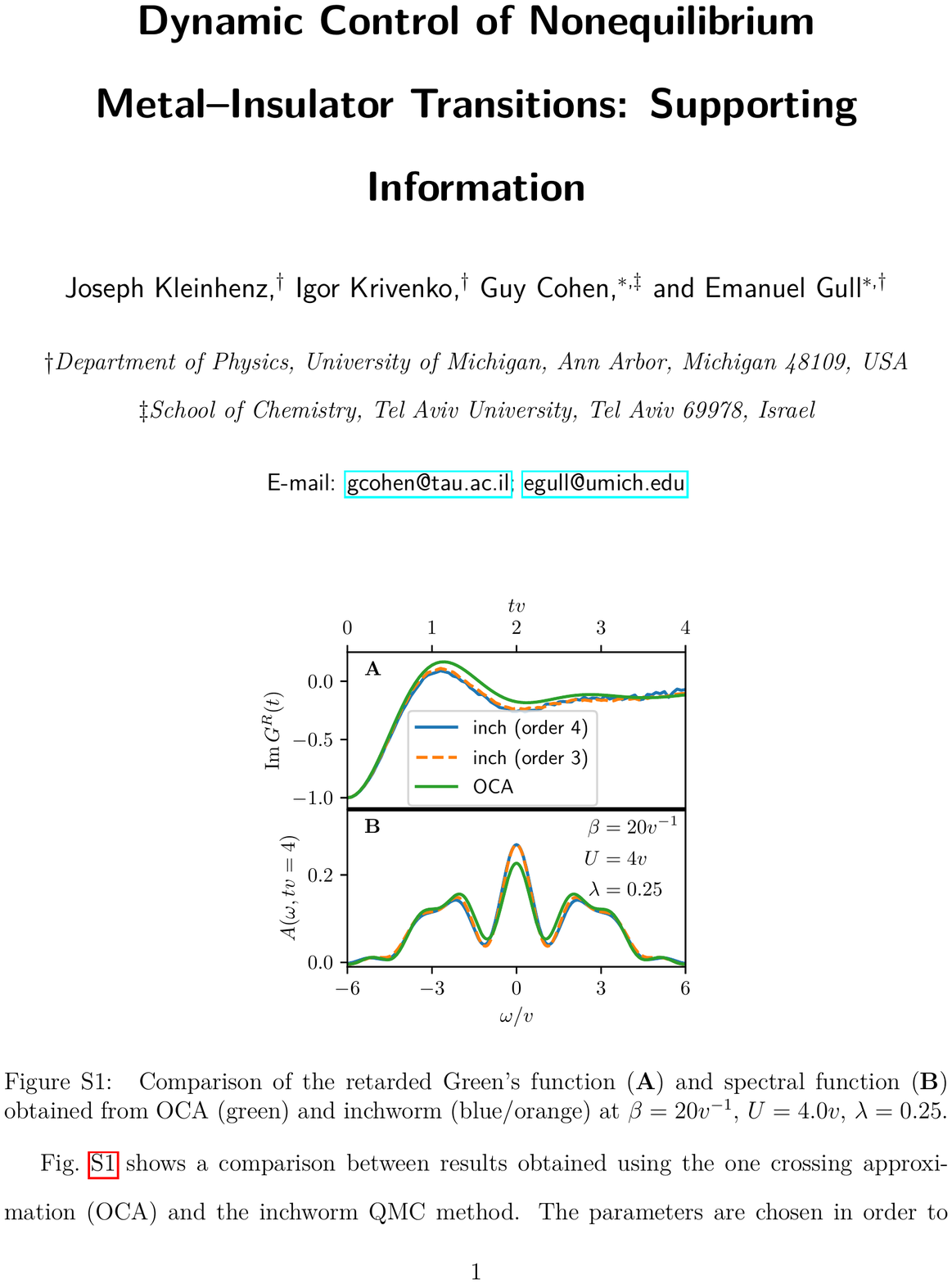}

\end{document}